\documentclass[conference]{IEEEtran}

\usepackage[bookmarks=false]{hyperref}
\usepackage{graphicx}% Include figure files
\usepackage{xcolor}
\usepackage{hyperref}

\def\code#1{\texttt{#1}}

\begin{document}
\bstctlcite{IEEEexample:BSTcontrol}
\title{sputniPIC: an Implicit Particle-in-Cell Code for Multi-GPU Systems}

\author{\IEEEauthorblockN{Steven W. D. Chien\IEEEauthorrefmark{1},
Jonas Nylund\IEEEauthorrefmark{1},
Gabriel Bengtsson\IEEEauthorrefmark{1},
Ivy B. Peng\IEEEauthorrefmark{2},
Artur Podobas\IEEEauthorrefmark{1} and
Stefano Markidis\IEEEauthorrefmark{1}}
\IEEEauthorblockA{\IEEEauthorrefmark{1}\textit{Division of Computational Science and Technology}, KTH Royal Institute of Technology, Stockholm, Sweden}
\IEEEauthorblockA{\IEEEauthorrefmark{2}\textit{Lawrence Livermore National Laboratory}, Livermore, CA, USA\\
}
\IEEEauthorrefmark{1}\{wdchien,jonasnyl,gabben,podobas,markidis\}@kth.se,
\IEEEauthorrefmark{2}peng8@llnl.gov,
}

\maketitle

\begin{abstract}
	Large-scale simulations of plasmas are essential for advancing our understanding of fusion devices, space, and astrophysical systems. Particle-in-Cell (PIC) codes have demonstrated their success in simulating numerous plasma phenomena on HPC systems. Today, flagship supercomputers feature multiple GPUs per compute node to achieve unprecedented computing power at high power efficiency. PIC codes require new algorithm design and implementation for exploiting such accelerated platforms. In this work, we design and optimize a three-dimensional implicit PIC code, called \emph{sputniPIC}, to run on a general multi-GPU compute node. We introduce a particle decomposition data layout, in contrast to domain decomposition on CPU-based implementations, to use particle batches for overlapping communication and computation on GPUs. sputniPIC also natively supports different precision representations to achieve speed up on hardware that supports reduced precision. We validate sputniPIC through the well-known GEM challenge and provide performance analysis. We test sputniPIC on three multi-GPU platforms and report a 200-800x performance improvement with respect to the sputniPIC CPU OpenMP version performance. We show that reduced precision could further improve performance by $45\%$ to $80\%$ on the three platforms. Because of these performance improvements, on a single node with multiple GPUs, sputniPIC enables large-scale three-dimensional PIC simulations that were only possible using clusters.
	
\end{abstract}


% \fi
\IEEEpeerreviewmaketitle

\section{Introduction}
\label{sec:introduction}

Large-scale supercomputers and parallel codes have enabled unprecedented high-resolution and highly accurate plasma simulations of fusion devices~\cite{markidis2014signatures}, space and astrophysical systems~\cite{peng2015formation, peng2015energetic,peng2015kinetic,markidis2012collisionless}. Traditionally, codes for plasma simulations rely on MPI combined with OpenMP together with a data layout that can take advantage of the cache hierarchies. An example of such codes is iPIC3D~\cite{markidis2010multi}, a Particle-in-Cell (PIC) code for plasma simulations on supercomputers using an implicit discretization of governing equations, hence an implicit PIC code. iPIC3D targets large-scale parallel systems and has achieved a parallel efficiency of 80\% when running weak scaling tests up to million cores on the IBM Blue Gene/Q Mira at Argonne National Laboratory~\cite{markidis2016epigram}. Today, the largest supercomputers, such as Summit, Sierra, and Piz Daint, are all equipped with multiple GPUs per compute node. With the advent of GPUs on supercomputers, parallel PIC codes need to be re-designed to exploit these accelerators' computational power. %This work details the design and optimization principles for porting and optimizing an implicit PIC code to general multi-GPU compute node on HPC platforms.

In this paper, we introduce \emph{sputniPIC}, a new implicit three-dimensional PIC code that is designed and optimized to exploit the computational power of multi-GPU systems. The software takes its name from the fact that we mainly run sputniPIC for space plasma simulations. We use the same general algorithm of sputniPIC CPU-based counterpart iPIC3D, which provides massive parallelism in hybrid MPI and OpenMP~\cite{markidis2010multi}. While iPIC3D uses only double precision for floating-point operations, sputniPIC supports native use of single- and mixed-precision, to exploit the single-precision floating-point units on GPUs. We use a hybrid approach when designing the workflow, by executing solvers on the CPU, while computationally intensive workloads such as particle mover and interpolation are offloaded to available GPUs.  One significant design difference with iPIC3D is that we perform \emph{Particle Decomposition} for GPU threads instead of \emph{Domain Decomposition}. Furthermore, we exploit device-level parallelism by implementing asynchronous data movement using pinned memory and CUDA streams. To achieve this, we introduce a novel particle processing scheme, called \emph{Particle Batching}, taking inspiration from data sample batching used during the training of Deep Learning neural networks. Particle batching does not only enable overlapping between communication and computation but also enables calculations on data exceeding the size of available GPU memory. We summarize our contributions as the following:
% TODO check if particle batching is really novel
\begin{itemize}
	\item We design and implement an implicit PIC code: \emph{sputniPIC}, to exploit the computation power of a multi-GPU system.
	\item We detail the design strategy and optimization technique: \emph{Particle Batching} to achieve high-performance execution.
	\item We introduce native support of multiple-precision representations and achieved 45-80\% when using single precision, relative to using double precision on GPUs.
	\item We show that by running sputniPIC on a multi-GPU node, it is possible to achieve a comparable performance of its CPU counterpart iPIC3D when running on 4-8 nodes of a Cray XC40 supercomputer.
\end{itemize}

The paper is organized as follows. We first introduce the governing algorithms in implicit PIC codes in Section~\ref{pic}. Section~\ref{design} presents the design principles and optimization techniques of sputniPIC. We describe the experimental set-up and the simulation and performance results in Sections~\ref{experiment} and~\ref{sec:eval}.  We discuss previous work on implicit PIC and PIC porting to GPUs in Section~\ref{relwork}. Finally, we conclude the paper by discussing some limitations and future work on the topic.

\section{The Implicit Particle-in-Cell Method}
\label{pic}
The PIC method simulates plasma particles, such as electrons and protons, as computational particles by computing their trajectories. The forces between particles, e.g., Coulomb and Lorentz forces, are calculated using a mean-field defined on the nodes of a grid to avoid the direct calculation entailing $\mathcal{O}(N_p^2)$ operations, where $N_p$ is the number of particles. We calculate the fields on the grid points by solving Maxwell's equations given charge and current density on the grid. Details of different formulation of PIC methods are presented in the computational plasma physics textbooks~\cite{birdsall2018plasma, hockney1988computer}.

In the PIC method, after the initialization of particle positions, velocities, and electric and magnetic fields, three distinct stages are repeated at each simulation time step: 1. Particle Mover (also Particle Pusher), 2. Particle to Grid Interpolation (or Moment Calculation), 3. Field Solver.

{\bf 1. Particle Mover}. The particle mover phase solves the equation of motion for each computational particle with position $\mathbf{x}_p$ and a velocity $\mathbf{v}_p$. sputniPIC, exactly like iPIC3D, uses an implicit in-time discretization scheme for the particle equations of motion. To solve the implicit discretized particle equations of motion (here in CGS units), we use a predictor-corrector scheme for calculating the average velocity $\mathbf{\bar{v}}_p = (\mathbf{v}_p^{n} + \mathbf{v}_p^{n+1})/2$  during the time step $\Delta t$ with $n$ indicating the time level:
\begin{eqnarray}
\label{vhat2}
\tilde{\mathbf{v}}_p&=&\mathbf{v}_p^n+\frac{q\Delta t}{2m}\mathbf{\bar{E}}_p\\
\label{vn+1/2}
\mathbf{\bar{v}}_p&=&\frac{\tilde{\mathbf{v}}_p+\frac{q\Delta
		t}{2m c}\bigl(\tilde{\mathbf{v}}_p\times\mathbf{\bar{B}}_p+\frac{q\Delta
		t}{2m c}(\tilde{\mathbf{v}}_p\cdot\mathbf{\bar{B}}_p)\mathbf{\bar{B}}_p\bigr)}{(1+\frac{q^2\Delta t^2}{4m^2c^2}{\bar{B}_p}^2)},
\end{eqnarray}
where $p$ is the particle index, $q,m$ are the particle charge and mass, and $c$ is the speed of light in vacuum.
The number of iterations to determine $\mathbf{\bar{v}}_p$ is either set by a prescribed error tolerance or fixed to a small number of iterations. In this work, we use three iterations for both electron and proton particles.  The $\mathbf{\bar{v}}_p$ calculation requires the electric and magnetic field at the particle position, $\mathbf{E}_p$ and $\mathbf{B}_p$. However, the values of the electric and magnetic field values, $\mathbf{E}_g$ and $\mathbf{B}_g$ are only defined at the grid points in the PIC method. To calculate these values, the PIC method uses the interpolation (or weight) functions $W({\bf x}_g-{\bf x}_p)$ defined as follows:
\begin{equation}
\label{interpW}
W({\bf x}_g-{\bf x}_p) =
\left\{
\begin{array}{l}
1 - |{\bf x}_g-{\bf x}_p|/\Delta x \quad \textup{if} \quad |{\bf x}_g-{\bf x}_p| < \Delta x \\
0  \quad \textup{otherwise}  .\end{array}
\right.
\end{equation}
In this case, we use linear interpolation function but higher order interpolation functions can be used. With the usage of interpolation functions, we can calculate the electric and magnetic field at the particle position from these values on the grid point $g$:
\begin{equation}
{\bf E}_p=\sum_g^{N_g} {\bf E}_g W({\bf x}_g-{\bf x}_p) \quad \quad {\bf B}_p=\sum_g^{N_g} {\bf B}_g W({\bf x}_g-{\bf x}_p) .
\label{interp}
\end{equation}
Once the particle average velocity is calculated, each particle position and velocity is updated as follows:
\begin{equation}
\label{dif_eom2}
\left\{
\begin{array}{l}
\mathbf{v}_p^{n+1} = 2 \mathbf{\bar{v}}_p - \mathbf{v}_p^{n}  \\
\mathbf{x}_p^{n+1} = \mathbf{x}_p^{n} + \mathbf{\bar{v}}_p\Delta t.
\end{array} 
\right. 
\end{equation}
Detailed descriptions of mathematical derivation of sputniPIC discretized equations can be found in~\cite{markidis2014fluid,markidis2011energy}. 

An important point for this work is that the mover takes most of the computational time in PIC. While the actual percentage of time taken for the particle from the mover depends on the problem under study (the number of particles and given CPU and memory systems), the particle mover percentage generally varies between 68\% and 73\%~\cite{sishtla2019multi} in typical space simulations. GPU computing and particle decomposition strategy, instead of traditional domain decomposition, are effective approaches to speed up the particle mover step.

{\bf 2. Particle to Grid Interpolation.} In this stage, we calculate the quantities that are input or sources for the field solver. In the implicit PIC method, these quantities are charge density, $\rho_g$, current density, $\mathbf{J_g}$, and the pressure tensor density, $P_g$. These quantities are all defined on the grid points and are calculated from the particle positions and velocities. Similarly to the calculation of electric and magnetic fields at the particle position in the particle mover phase, we use the interpolation functions $W({\bf x}_g-{\bf x}_p)$ to determine $\rho_g, \mathbf{J}_g, P_g$ at the grid point $g$:
\begin{equation}
\{ \rho,\mathbf{J}, P \}_g  =  \sum_p^{N_p} q \{1,  \mathbf{v}_p, \mathbf{v}_p \otimes \mathbf{v}_p  \} W({\bf x}_g-{\bf x}_p) .
\end{equation}

The particle to grid interpolation is the second most computationally intensive part of the PIC method. In simulations of magnetic reconnection (an explosive phenomenon in Earth's magnetosphere), it typically takes approximately 25\% of the whole computational cycle~\cite{sishtla2019multi}. Therefore, to use GPUs also for this step improves the performance of the entire code.

{\bf 3. Field Solver}. The third and final step of the implicit PIC method is the solution of discretized Maxwell's equations on the grid. This step takes $\rho_g, \mathbf{J}_g, P_g$ as input and computes ${\bf E}_g$ and ${\bf B}_g$. The implicit PIC solves a linear system arising from the discretization of Maxwell's equations implicitly in time with a Generalized Minimal Residual (GMRes) linear solver. In addition to GMRes, we solve a discretized Poisson equation with the Conjugate Gradient (CG) at each computational cycle to ensure the continuity equation is satisfied~\cite{markidis2011energy}. This additional step is also called \emph{divergence cleaning}. In typical implicit PIC simulations, the linear systems solved with the GMRes takes 5-10 $\times$ the time to solve the Poisson equation. However, in implicit PIC simulations, the solver takes typically only 6\%. For this reason, in this work, the field solver is still executed on the CPU, and we do not take advantage of multi-GPU systems. 

\section{Design and Implementation}\label{design}
\begin{figure}[tb]
\begin{center}
	\includegraphics[width=1\columnwidth]{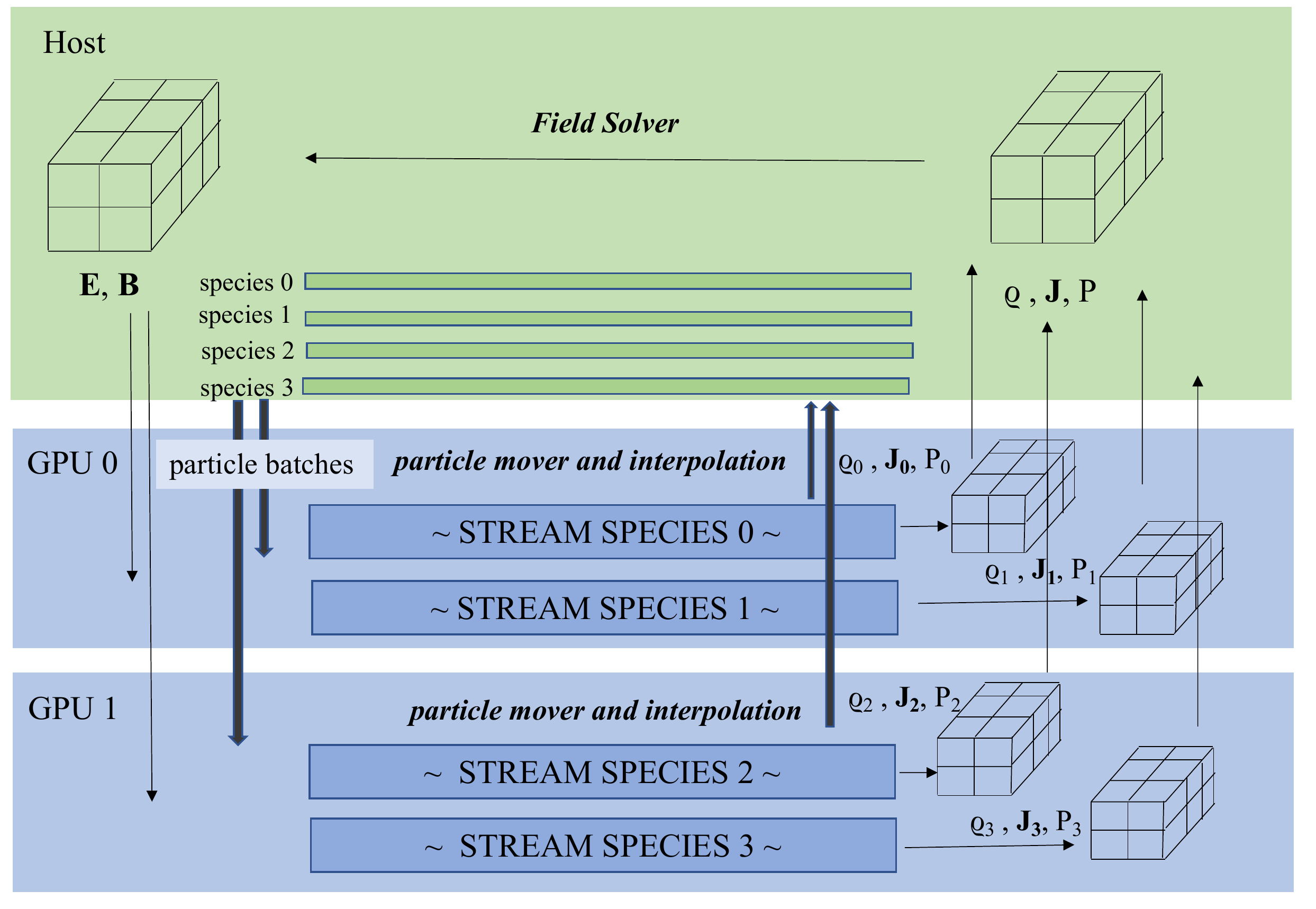}
	\caption{An overview of the sputniPIC workflow. $N_s$ particles species ($N_s=4$ in the example) are split into $M$ particle batches. Particle batches are distributed over the available GPUs. GPUs compute moments $\rho_s, \mathbf{J}_s, \mathbf{P}_s$ for species $s$ from each particle batch and then stream to the host. The host updates the fields ({\bf E}, {\bf B}). Updated fields are then used by GPUs to update particle positions.\label{fig:sputnipic-workflow}}
	\end{center}
\end{figure}

We design sputniPIC specifically for HPC systems accelerated with multiple GPUs per node. sputniPIC leverages GPU for the compute-intensive particle mover and interpolation while solving fields on CPU. Figure~\ref{fig:sputnipic-workflow} illustrates the main workflow. Particles are separated into species, e.g., species 0-3. For each species, sputniPIC launches an independent stream on GPU for particle mover and moments interpolation. These streams are distributed over all the available GPUs. At each time step, GPUs interpolate moments from all species, and then the host uses the moments, i.e., $\rho$, $\mathbf{J}$, and $P$, for solving electric and magnetic fields. Once these fields are updated, i.e., $\mathbf{E}$ and $\mathbf{B}$ in Fig.~\ref{fig:sputnipic-workflow}, they are transferred to the GPUs to update particle positions and velocities.

\subsection{Particle Decomposition}
We design a particle decomposition scheme in sputniPIC. Typical PIC simulations use billions of particles that dominate the computation time and memory footprint. The distribution of these particles in a simulation box can be highly skewed depending on the physics phenomena. Therefore, domain decomposition, a common design of HPC applications on distributed systems, can result in a significant load imbalance on GPUs. Also, if the simulation box is decomposed into multiple subdomains, when a particle moves from one subdomain to another, particle communication becomes necessary. This communication is dynamic and irregular as the number of particles moving out of a subdomain is only known at runtime. 

Our design decomposes particles to multiple GPUs. Each GPU is assigned species in a round-robin fashion. As species typically have similar particle count, this results in a fair share of particles per GPU. Each GPU also keeps a private copy of fields of the whole simulation box, including moments and electric and magnetic fields. Note that even fields of the entire simulation box require a much lower memory footprint than that of particles. During the particle interpolation, each GPU deposits the moments from their particles to their local copy of fields. These fields of moments, i.e., $\rho$, J, and P, from all GPUs, are then consolidated on the host for updating the electric and magnetic fields. 

One challenge in the particle decomposition is the reduced data locality of the field data structure when particles start moving in the simulation box as the simulation evolves. At the beginning of the simulation, particles assigned to each GPU mostly reside in proximity in the simulation box. Therefore, interpolating their moments to the field will likely access the same subdomain of the field, exhibiting a good cache locality on the field data structure. As the simulation evolves, particles start moving in the simulation box so that interpolation will likely update to different subdomains of the fields, resulting in low data locality. We address this challenge by particle sorting and re-decomposition. sputniPIC triggers particle sorting to order particles by their position in the simulation and then re-assigns particles in proximity in the simulation box to GPU. This procedure is typically infrequent and improves data locality in particle interpolation and mover.

\subsection{Particle Batching and Pipelining}
We design a particle batching scheme to improve communication and computation overlapping and enable large simulations beyond the GPU memory capacity. Each PIC simulation may use multiple particle species, such as electron and protons, with different charges and mass ratios. These species could have different particle populations, e.g., electrons in the background with no drift velocity or electrons with a drift velocity forming a current. sputniPIC separates different particle species into separate streams of operations to maximize throughput. Each stream performs interpolation and mover independently on different particle species without data dependency.  

Each particle species may have a large number of particles that exceed the GPU memory capacity. sputniPIC further divides each particle species into a finer granularity, called \textit{particle batches}. Particle batching improves the pipelining of communication and computation within one particle species. Also, particle batches from different particle species could overlap with each other. Data transfer for one particle batch can be effectively overlapped with the computation of another batch. If not all particles can fit in the GPU memory, sputniPIC inserts a swapping operation to swap out particle batches before bringing new batches for computation on the GPU. The batch size is a configurable parameter in sputniPIC because its optimal value depends on the underlying hardware and simulation setup. 

sputniPIC fuses particle interpolation and mover into one pass to improve data reuse in the cache and reduce kernel launching overhead. As we discussed before, PIC codes consist of major steps of particle interpolation, field solver, and then particle mover, where the output from a step is used as input for the next one. In the particle interpolation, all particles are iterated to deposit their charges and moments to the field. In the particle mover, all particles are iterated to update their positions by the field forces. By fusing these two steps, each particle only needs to be fetched into the cache once, and data reuse can be improved. Therefore, sputniPIC merges the two steps into one kernel such that once the position of a particle is updated, the new position is used for interpolating particle charges and moments to the field immediately. This single-pass particle computation significantly reduces the data movement on GPU and kernel launching overhead. 

\subsection{Multi-precision Support}
sputniPIC adopts normalized units in simulations and leverages the statistical nature of PIC codes to tolerate low-precision calculations. PIC simulations are inherently noisy since they rely on a statistical description of the plasma by using particle distribution functions. Massive amounts of particles are used in a simulation to reconstruct statistically representative distribution. However, this also causes the particle data structure to dominate the memory footprint. Modern GPU hardware supports fast and power-efficient low-precision arithmetics. For instance, Nvidia Volta V100 GPUs have 64 single-precision ALU but only 32 double-precision FPUs per Streaming Multiprocessor~\cite{markidis2018nvidia}. Naturally, changing the particle data structure from double to single precision could directly halve the memory footprint and data movement on GPU and speed up the computation. 

sputniPIC features a particle data structure to support different precision requirements. The users are provided with the flexibility to select the precision format based on the simulation setup. One challenge of directly replacing double-precision floating-point format with lower precision formats is the possibility of leading to large rounding errors. To address this problem, a user can choose to employ a mixed-precision approach: double-precision for field data structure and field computation, while keeping single-precision on the GPU for the particle mover and interpolation quantities. In other words, sputniPIC achieves mixed-precision compute by using different precision in different parts of the PIC cycle, where incoming data are cast as appropriate. 

\subsection{Implementation} 
We implement sputniPIC as a C++/CUDA code with OpenMP parallelization on the host. sputniPIC features the Structure of Array (SoA) data layout to simplify data transfer to GPU. Unlike the Array of Structure (AoS) data layout, no temporary copies or staging are required before transferring data between GPU and CPU. The particle data structures can be implemented in either single or double precision. Similarly, the field data structures can be in single or double precision floating operations, with a private copy on each GPU and the host. 

sputniPIC uses pinned host memory to avoid extra data copy from the host pageable memory to the pinned host array. CUDA performs Direct Memory Access (DMA) through PCI-E or NVLink to transfer data between the device and host. However, DMA cannot directly access the host pageable memory region so that data in this address space must be copied to a staging area before DMA transfers. sputniPIC allocates data structures that need to be communicated between host and GPU in the pinned host memory directly to avoid this extra data copy.  

sputniPIC takes advantage of multiple CUDA streams to implement the particle batching and pipelining strategy. These streams are distributed over all the available GPUs to exploit multi-GPU systems. The performance of particle computation scales up almost linearly with the number of used GPUs. Asynchronous memory communication with \code{cudaMemcpyAsync()} is used for data transfer in the CUDA streams.

\section{Experimental Environment}
\label{experiment}
We compiled sputniPIC using CUDA 10.1 and evaluated its performance on three platforms. The architecture of each system is summarized as follows:

\begin{itemize}
	\item {\bf 2xP100 + Intel} is a GPU node on Flash cluster at Livermore Computing. The node consists of an Intel Xeon E5-2670 processor with 256 GB DRAM and two Nvidia Tesla P100 GPUs with 16 GB HBM each. CPU and GPU are interconnected by PCIe links. The node runs RHEL 7.7 and the host compiler is GCC 8.1.
	\item {\bf 2xV100 + Intel}  is a GPU node on Kebnekaise at HPC2N in Ume\.a. It has two Intel Xeon Gold 6132 processor ($2\times14$ cores) with 192 GB RAM. The node has two Volta V100 GPUs with 16 GB memory and the GPU is connected through PCIe. The operating system is Ubuntu 16.04 and the host compiler is GCC 8.3.
	\item {\bf 4xV100 + Power9} is a node on the unclassified Sierra system, Lassen cluster. Each node has an IBM Power9 processor with 256 GB DRAM and four Volta V100 GPUs. Each GPU has 16 GB HBM2 device memory. The GPUs are interconnected through NVLINK2 to CPU. The system runs REHL 7.6 and GCC 8.3.
\end{itemize}

We choose a well-known simulation challenge in space physics -- the GEM challenge~\cite{birn2001geospace}  -- for simulating the magnetic reconnection phenomenon. The simulation used parameters that are derived from observations of the Earth magnetotail. Our simulation uses a three-dimensional domain box that is more realistic than the simplified two-dimensional configuration in the original GEM challenge. Furthermore, our simulation features a higher charge-to-mass ratio, 64, instead of 25, to mimic a realistic ion-to-electron mass ratio. The grid consists of $128 \times 64 \times 64$ cells and four particle species are in use. Each particle species is initialized with 125 particles per cell. The total number of particles is approximately $2.6E8$. For the performance evaluation, we advance the simulation for 100 time steps, with each step equal to $\omega_{pi} \Delta t = 0.25$, where $\omega_{pi}$ is the ion plasma frequency. We also perform a complete simulation of the GEM challenge in 3,000 computational cycles and present the results of this simulation in Section~\ref{sec:eval}. 

We compare the performance of sputniPIC in single-precision and double-precision on GPUs and also on CPU-only baseline that is parallelized with OpenMP. When using OpenMP, we set the number of threads to be equal to the number of cores on a compute node. To have a comparison with iPIC3D, we perform the same performance tests running iPIC3D on a Cray XC40 supercomputer (called Beskow) that has not GPU. Each Beskow node features two 16-cores Xeon E5-2698v3 Haswell processors. We compare sputniPIC performance with that of iPIC3D running on one, two, four, and eight Beskow nodes. In this work, we choose the optimal configuration (Section~\ref{sec:batch}) on the 2xV100 + Intel system for all the tests on the three different platforms: 256 Threads per Block (TPB) are used, together with 16 particle batches per species.

\begin{figure}[tb]
	\begin{center}
		\includegraphics[width=1\columnwidth]{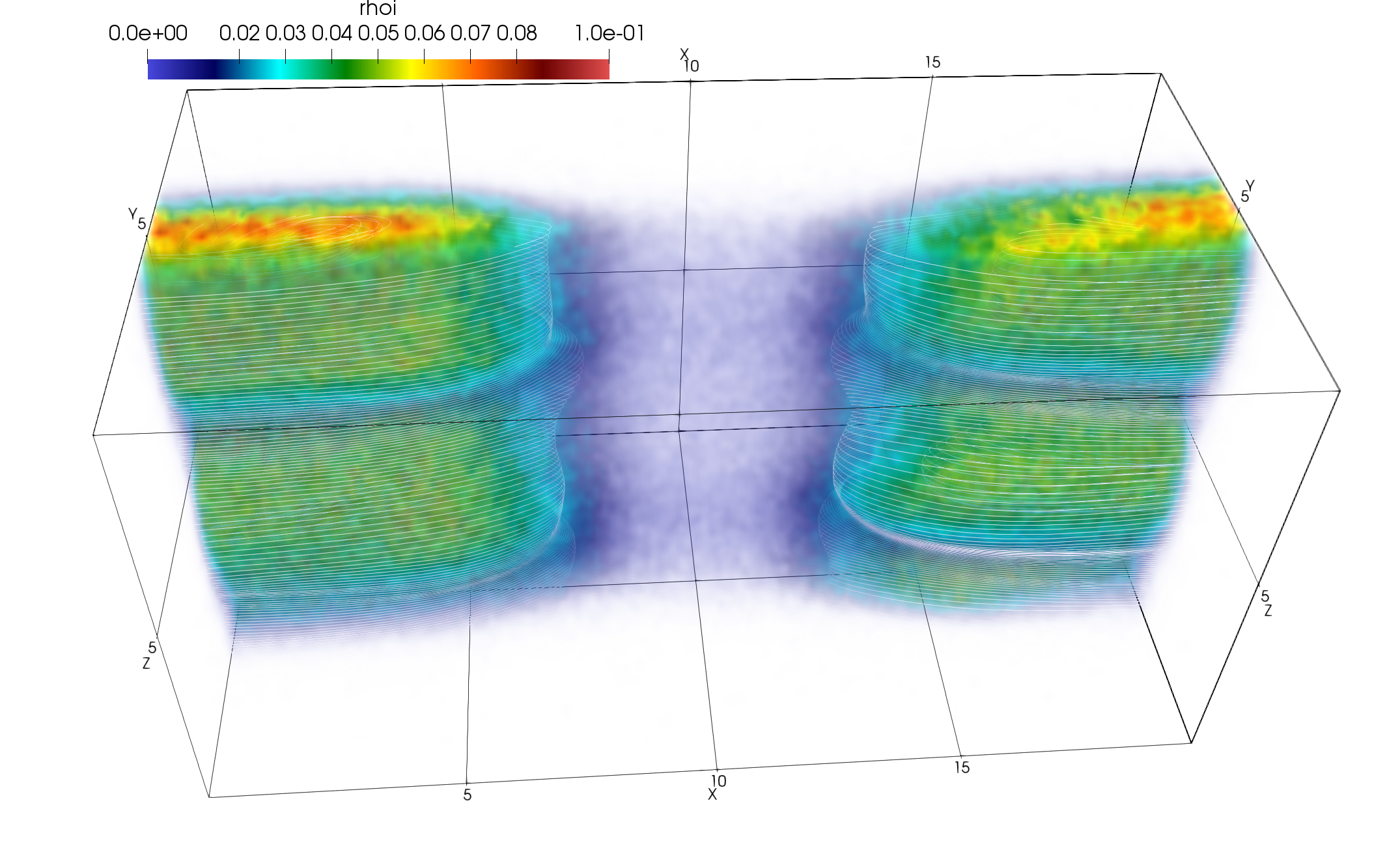}
		\caption{Iso-surfaces of ion charge density with superimposed magnetic field lines after 3,000 computational cycles in a three-dimensional sputniPIC simulation of the GEM challenge test.\label{picresult}}
	\end{center}
\end{figure}

The performance comparison uses Millions of Particles Advanced and interpolated per second \textbf(MPA/s) as the figure of merit reported in the result section. We calculate MPA/s from dividing the total number of particles in the simulation by the time spent in the particle mover and interpolation per computational cycle. We use MPA/s as the performance indicator because the particle mover and interpolation phases dominate the execution time. We report the average over 100 cycles together with the standard deviation to show the performance variability.

\section{Evaluation}
\label{sec:eval}
In this section, we first present the simulation results of the GEM challenge, and then we show the sputniPIC performance results.

\subsection{Simulation Results}
Our first test is to verify that sputniPIC produces the correct simulation results with a physical meaning. We run the GEM challenge test for 3,000 cycles and verify that magnetic reconnection occurs correctly with the generation of plasma jets and the reorganization of the magnetic field topology. Visually, Fig.~\ref{picresult} shows the correct and expected behavior: we can observe the expected formation of plasma jets represented by iso-surface of ion charge density, $\rho_i$. We can also observe the formation of a magnetic field island represented by the magnetic field lines (white lines). 

%Our first test is to verify that sputniPIC produces the correct simulation results with a physical meaning. We run the GEM challenge test for 3,000 cycles and verify that magnetic reconnection occurs correctly with the generation of plasma jets and the reorganization of the magnetic field topology. Figure \ref{picresult} shows the formation of plasma jets that represented by the iso-surfaces of electron charge density, $\rho_e$, and the formation of a magnetic field island represented by the magnetic field lines (grey tubes).

\subsection{Overall Performance}
%Figure \ref{fig:sputnipic-gpu-performance} shows the performance of particle mover and interpolation on the four platforms.
\begin{figure}[t]
	\begin{center}
		\includegraphics[width=1.0\columnwidth]{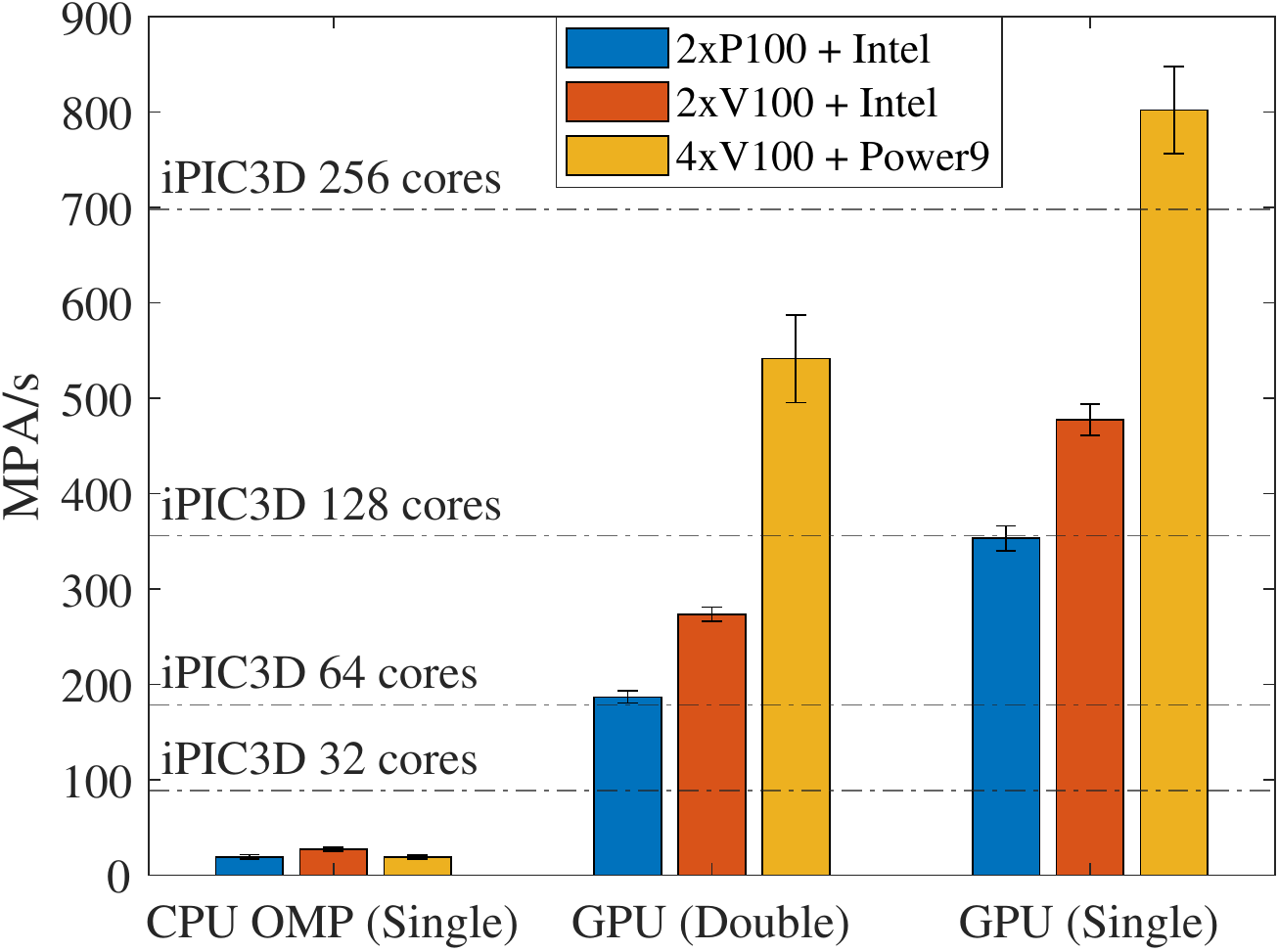}
		\caption{Performance of sputniPIC particle mover and interpolation on our multi-GPU platforms. The corresponding performance of iPIC3D on Beskow supercomputer is plotted as lines for comparison. The relatively low performance of sputniPIC CPU OMP to iPIC3D is because of the interpolation step that cannot be completely parallelized. On the GPU this step is implemented with CUDA atomics.}
		\label{fig:sputnipic-gpu-performance}
	\end{center}
\end{figure}

In this section, we compare the performance that we empirically measured on our three systems running the sputniPIC particle movers and interpolation, and position that performance to the original iPIC3D code in order to reason around the gains of specializing our particle-in-cell method towards modern GPUs. Fig.~\ref{fig:sputnipic-gpu-performance} shows the performance for our different systems.

We start by observing that the CPU-only version of sputniPIC parallelized with OpenMP experience significantly lower performance (less than 30 MPA/s) than those that involve GPUs; we can observe nearly a magnitude of performance difference between sputniPIC on CPUs compared to sputniPIC on GPUs. We also note that the performance is fairly uniform across the different CPU architectures, leading to similar performance profiles for sputniPIC running on both Intel and IBM Power9 processors.

The GPU-accelerated versions running on NVIDIA P100 and V100 are significantly faster than the OpenMP versions, and experience between 180 MPA/s and 800 MPA/s, depending on the type, the number of accelerators, and the numerical representation. We observe that both the NVIDIA P100 and V100 GPU both experience a significant speedup (up-to approximately 90\% increase) when going to single-precision. Overall, we note that for sputniPIC, relaxing the numerical representation is a viable way of increasing GPU performance.  We also note that the performance increase experienced by sputniPIC when moving across two generations of NVIDIA GPUs (P100 and V100) is 35\% and 46\% for single-precision and double-precision respectively. 

When we position sputniPIC to its parent, the iPIC3D code running on the Beskow supercomputer, we note several advantages of using GPUs. The performance reached on the dual NVIDIA P100 systems is slightly higher than that of two nodes of Beskow (64 cores), both reaching ~180 MPA/s (186 MPA/s for 2xP100 and 178 MPA/s for two Beskow nodes). A similar performance increase can be seen for the two and four NVIDIA V100 systems, which can deliver performance equivalent to three Beskow nodes (2xV100) and four Beskow nodes (4xV100) of performance.  If allowed to run using single-precision arithmetics, our largest GPU-node (4xV100) would reach an impressive 0.8 GPA/s, which would be faster than eight Beskow nodes of iPIC3D (which runs double-precision). 

Overall, we can conclude that the addition of GPUs for PIC codes, such as sputniPIC, and the specialization of PIC methods to leverage GPUs can significantly boost single node performance and provide comparable performance to that of multiple nodes of existing supercomputers.

\subsection{Impact of Particle Batch and Thread Block}
\label{sec:batch}
\begin{figure}[t]
	\begin{center}
		\includegraphics[width=\columnwidth]{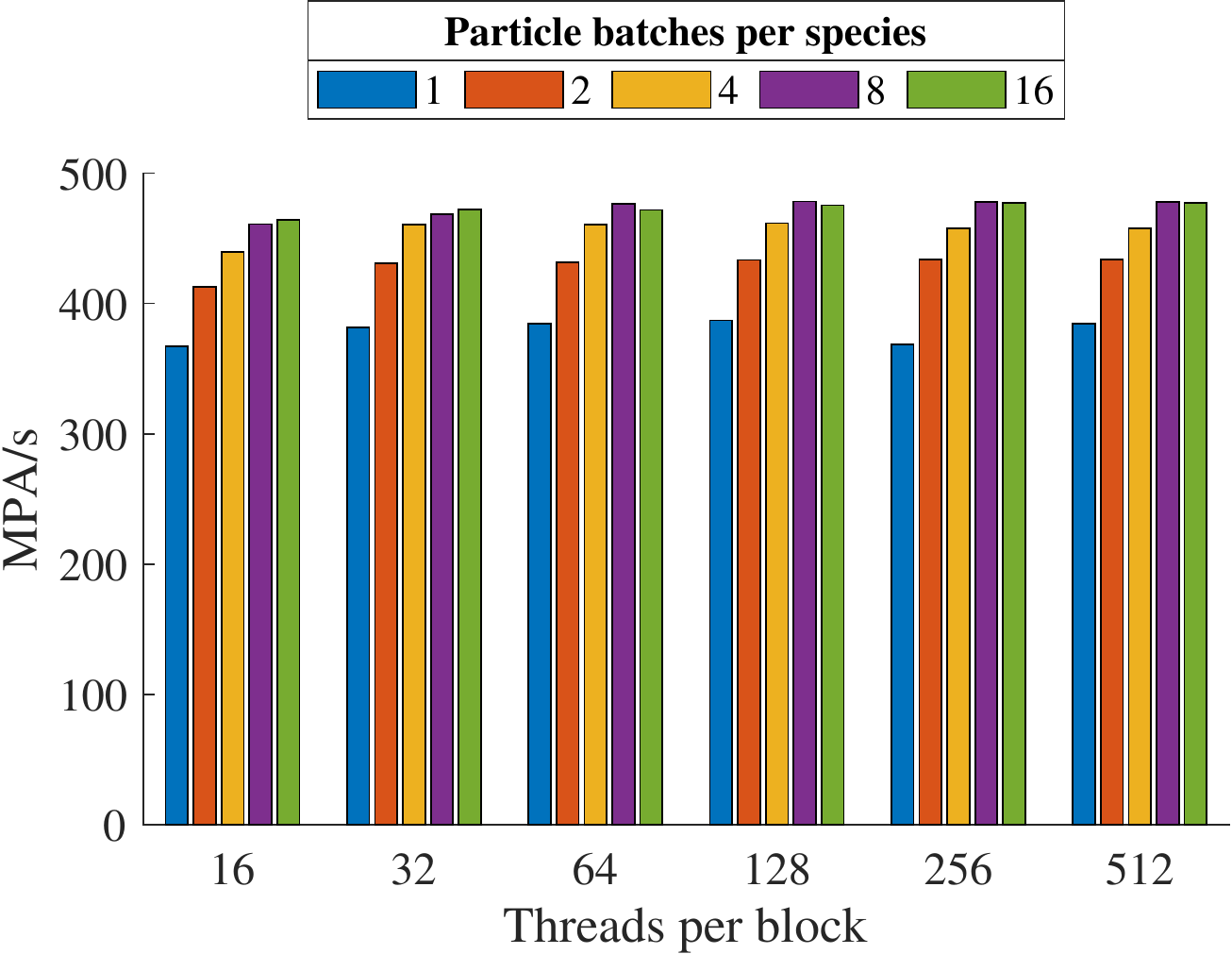}
		\caption{Impact of performance using a different number of particle batches per species and threads per block on the 2xV100 + Intel system. We do not report the error bars as the standard deviation is less than 5\%.}
		\label{fig:sputnipic-numerofbatches-TPB}
	\end{center}
\end{figure}

The optimal number of particle batches and TPB depends on the platform. By investigating Fig.~\ref{fig:sputnipic-numerofbatches-TPB} (performance on the 2xV100 + Intel system), it is clear that performance varies slightly with the number of TPB. The CUDA block size does not have a significant impact on the results, varying by just a few percent, which is within the margin of error for the measurement. On the other hand, the number of batches had a larger impact: an increased number of batches, e.g., fewer particles per batch, improves the performance up to 16 particle batches per species, where the performance gain leveled off. Updating the particles in 16 batches instead of one batch provides a performance increase between 20-25\%. 

\subsection{Multi-precision Compute}

\begin{figure}[t!]
	\begin{center}
		\includegraphics[width=\columnwidth]{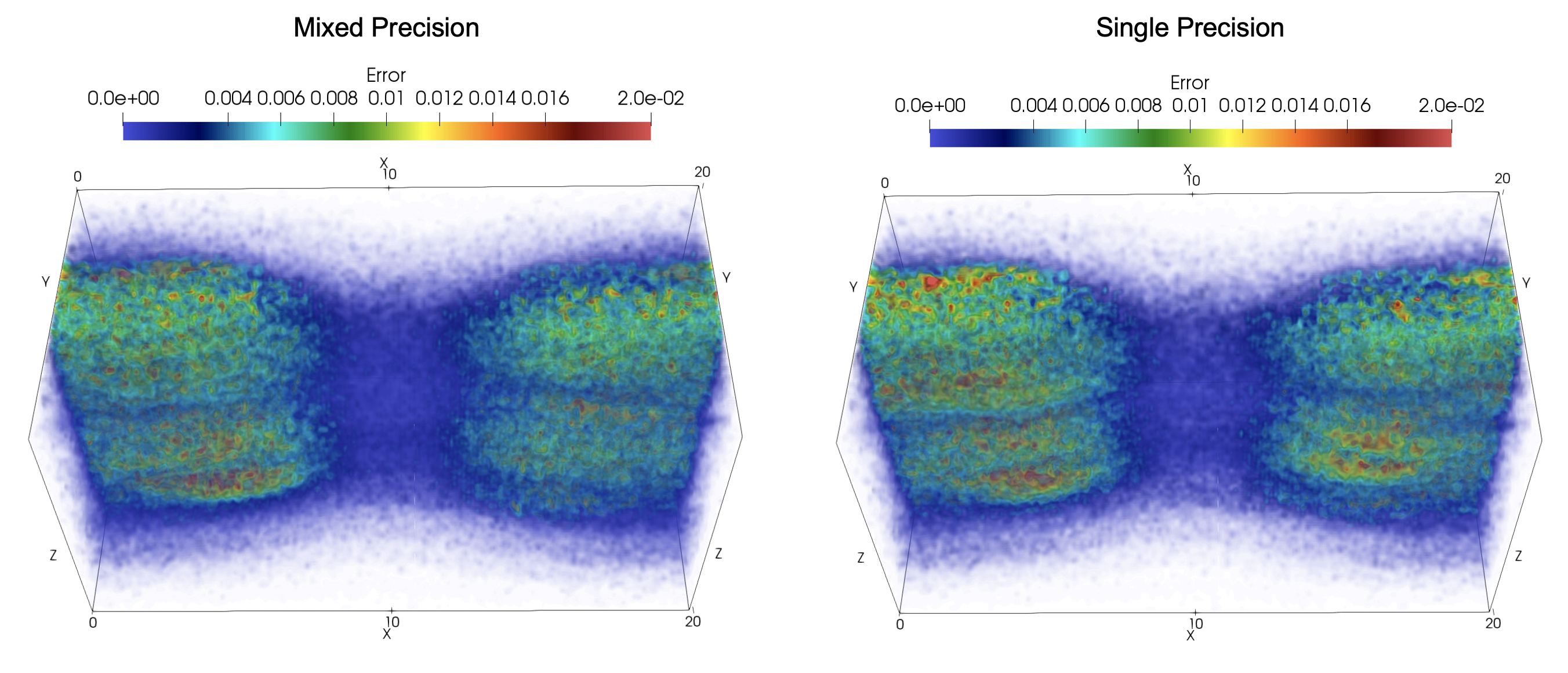}
		\caption{Absolute error by data point-wise comparison in $rho_e$, against the same sputniPIC simulation in double-precision. Using mixed-precision (left) gives improved accuracy over single-precision (right).}
		\label{fig:mixed-precision-error}
	\end{center}
\end{figure}

One important feature of sputniPIC is the native support of single precision. Furthermore, it is possible to use mixed-precision, where the GPU workloads are computed in single-precision with CPU workloads in double-precision. By switching the entire simulation from double to single precision, we observe an increase in error on all products. For instance, $rho_e$ and $rho_i$ give an error norm of 2.8506 and 2.9637 respectively. However, once mixed-precision is used, they reduced to 2.5379 (-10.97\%) 2.6451 (-10.75\%). At the same time, there is little to no impact on the error on the fields, since their input data are computed in single-precision. The error data point-wise error of $rho_e$ is visualized in Fig.~\ref{fig:mixed-precision-error} where the product from single-precision runs gives a larger area with redness (higher in error). We note that mixed-precision has very little impact on the GPU kernel's performance, where the overhead of value casting is introduced. 

\subsection{Computation and Communication Overlapping}
\begin{figure*}
	\begin{center}
		\includegraphics[width=1\textwidth]{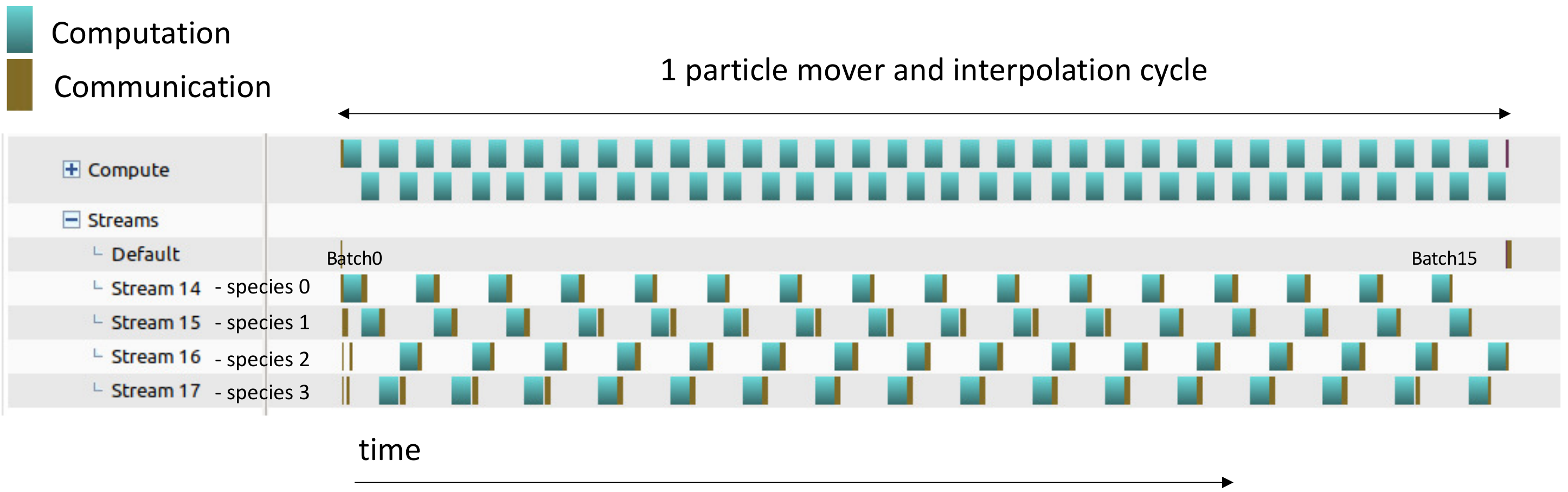}
		\caption{Profiling results of the particle mover and interpolation stages show effective overlapping between kernel executions and memory operations.}
		\label{fig:nvvpScreen}
	\end{center}
\end{figure*}

The performance of particle mover and interpolation relies on efficient overlapping between the communication and computation in time when the particles are batched onto GPU. Fig.~\ref{fig:nvvpScreen} shows the profiling information from sputniPIC during one iteration of the particle mover and interpolation using the Nvidia Visual Profiler. We use four particle species and, therefore four streams. 16 particle batches per species are used in this test.

\section{Discussion}
In this work, we designed and implemented an implicit PIC code specifically for compute nodes with multiple GPUs. The new sputniPIC code uses the same numerical algorithm and basic building blocks of the iPIC3D code~\cite{markidis2010multi}. One main contribution in sputniPIC is the novel particle decomposition design, instead of the domain decomposition in iPIC3D, to effectively utilize the massive parallelism on multi-GPU accelerated platforms. Another contribution in sputniPIC is the native support for different precision representation that could take advantage of modern GPU single-precision compute to further accelerate physics simulation. Finally, sputniPIC also proposes a convenient data layout of particle information for multiple GPUs. For the implementation, we achieve efficient overlapping between computation and communication by using CUDA pinned memory, streams, and asynchronous memory transfers. The particle mover and interpolation phases show a considerable performance improvement on all three platforms. 

After the improvement of 200-800x in the particle mover and interpolation with respect to the performance without GPU, the field solver is now the performance bottleneck in all the three platforms. On the platform, the field solver is now 6x slower than the fused mover and interpolation. Our current task is to port the field solver or part of it (the residual calculation at each iteration) to GPU.

\begin{figure}[t]
	\begin{center}
		\includegraphics[width=\columnwidth]{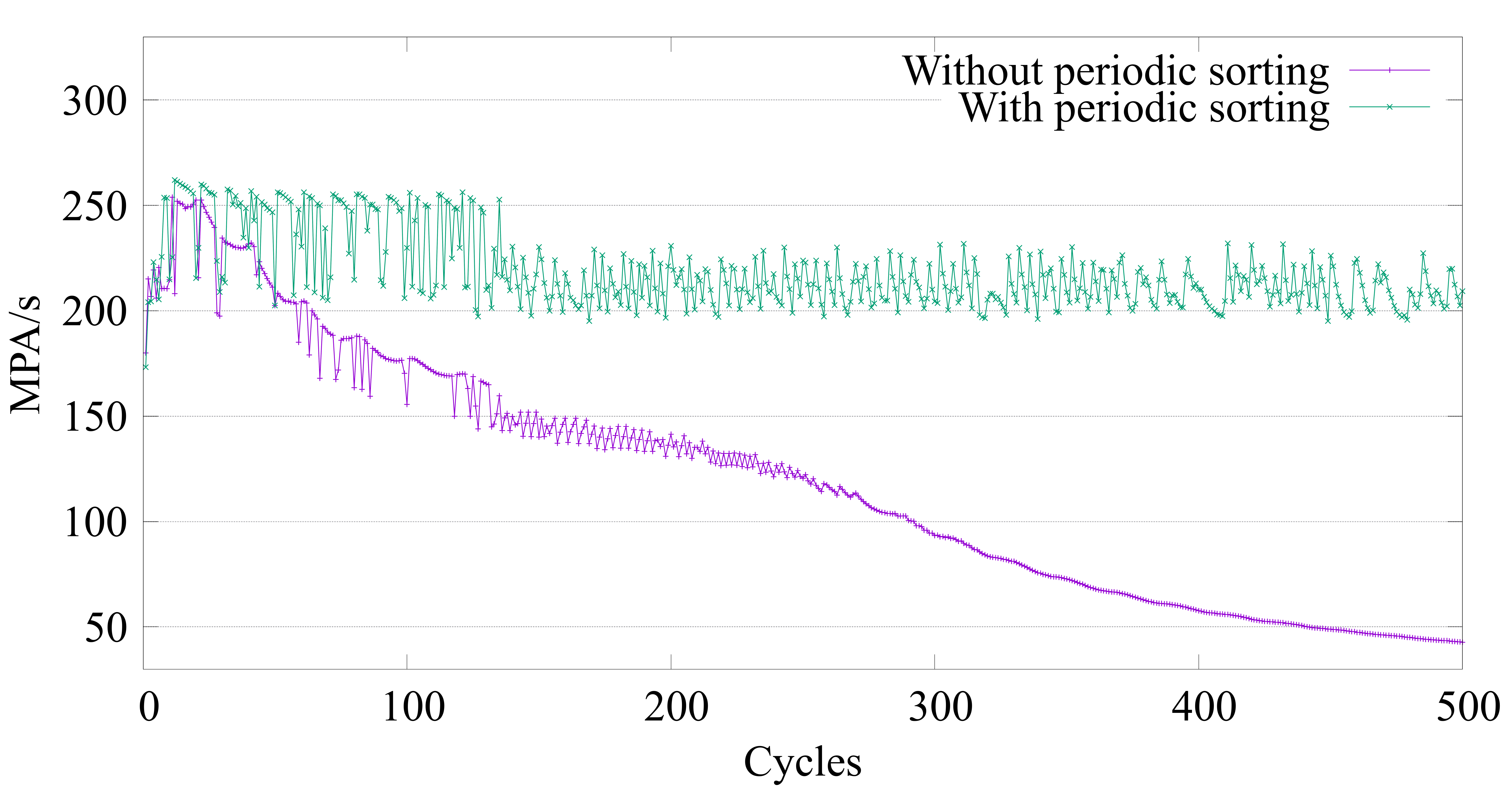}
		\caption{An experimental simulation for 500 cycles with and without particle sorting. The run with sorting is able to maintain high computation rate.}
		\label{fig:time-vs-cycles}
	\end{center}
\end{figure}

During a long simulation of 3,000 cycles, we noticed the performance of the GPU execution decreases over time. We investigate this on a local development platform through the CUDA Visual Profiler. We observe a sharp decrease in L2 cache hits together with a reduction in performance over time. While the benchmark runs in our performance results are relatively short and are not impacted heavily, this can seriously hamper long-running simulations. For this reason, we experimented with particle sorting in Fig.~\ref{fig:time-vs-cycles} on our local development platform with a 500 cycles simulation, with a tunable sorting period of 10 cycles, versus when no sorting is applied. It can be clearly observed that by invoking particle sorting, we are able to maintain a high and consistent performance of over 200~MPA/s over time, while the simulation without sorting quickly drops to 50~MPA/s at cycle 500.

%During long simulation over 3,000 cycles, we noted that the performance of particle mover and interpolation on GPU decreases in time. On our 2xV100+Intel platform, the performance decreased over a few thousand cycles and leveled off at about a quarter the initial performance, see Fig.~\ref{fig:time-vs-cycles}. Through CUDA visual profiler (nvvp), a sharp decrease in L2 cache hits could be seen at the same time, where cache hits decrease by almost 50\% in the first 100 cycles, and then continue decreasing. To mitigate this problem in long sputniPIC simulations, we implemented a particle sorting according to their location~\cite{nakashima2017large}.

A limitation of this work is that the current sputniPIC implementation does not support simulations on multiple nodes with GPUs. We are currently investigating the possibility of combining particle and domain decompositions to extend sputniPIC to use multi-node systems with MPI.

One of the most important results of this paper is that this work shows that an application specifically designed for modern multi-GPU systems could achieve performance that is comparable to the performance of highly tuned MPI code running on 4-8 nodes of a supercomputer without GPUs. This allows us to perform large-scale simulations, previously only possible on supercomputers, on a single node with multiple GPUs.

\section{Related Work}
\label{relwork}
The PIC method is one of the main computational tools for the simulation of space, astrophysical, and fusion plasmas~\cite{birdsall2018plasma}. The PIC method was initially developed in the late Fifties and early Sixties and then further improved by using more sophisticated numerical schemes, such as semi-implicit and fully-implicit schemes~\cite{markidis2011energy}, and combining fluid and kinetic equations for plasmas~\cite{markidis2014fluid}. In this work, we use the iPIC3D code algorithm that was initially developed at Los Alamos National Laboratory in 2005.  During the last decades, the code has been improved by using advanced algorithmic parallelization strategies and optimized I/O~\cite{markidis2016epigram,narasimhamurthy2018sage}. 

The PIC method is conveniently suited to exploit GPUs because of the particle mover can be easily expressed as vector operations. Several studies have focused on developing PIC codes specifically for GPU systems. The first seminal work on PIC porting to GPU systems is by Stantchev et al.~\cite{stantchev2008fast}: in particular, the paper presents introduces an optimized particle-to-grid interpolation. Optimization of the data layout in Fortran PIC codes for GPUs is presented in Refs.~\cite{decyk2011adaptable,decyk2014particle}. Widely-used code, such as WarpX~\cite{thevenetwarpx}, Osiris~\cite{fonseca2018osiris} and VPIC ~\cite{bird2019vpic}. However, all these previous works and PIC codes use an explicit in-time discretization of the PIC algorithm, and porting of implicit PIC method does not exist in the literature.

\section{Conclusions and Future Work}
\label{con}
Current state-of-the-art supercomputers feature multiple GPUs on each compute node to achieve high-performance computation with low power consumption. In order to exploit their computational power, existing software needs to be redesigned to adapt to the new programming and execution model. To enable fast PIC simulations on multiple GPUs, we introduced sputniPIC, an implicit PIC code that uses GPU friendly data layout and provides native support of multiple-precision in computation. The code takes a hybrid approach where the solvers are executed on the CPU and particle data computation is offloaded to the GPUs. Furthermore, we implemented a \emph{Particle Batching} scheme for batched particle computation, similar to the use of batched training of Deep Learning networks. Particle batching not only allows the computation of particle data that does not fit into GPU memory but also exploits asynchronous data movement to overlap with computation. We evaluated the correctness of the output products through a well-know GEM challenge and provided an in-depth analysis of the performance impact of particle batching, thread block sizes, and precision. Through performance testing, we showed that sputniPIC running with single precision on multi-GPU nodes can achieve a comparable performance of its CPU counterpart, iPIC3D when running on four to eight nodes of a Cray XC40 supercomputer.

One performance issue in sputniPIC is the reduction in particle mover and interpolation performance when the computation cycle increases. A reason, according to the profiler, is due to a reduction in the L2 cache hit rate. To improve cache efficiency, we implemented a particle sorting scheme that improves cache efficiency by periodically sorting particles according to their cell location.

Currently, sputniPIC supports execution on a multi-GPU node and a major limitation is that it does not support multi-node execution. While being able to provide a comparable performance of multi-node execution on CPU, this limits the scalability when very large simulations are performed. For this reason, we are currently extending our particle decomposition scheme to support \emph{Hierarchical Particle Batching}, where particles are first batched across computing nodes through MPI and further batched for GPUs within the nodes. We aim to introduce more advanced features and novel techniques in sputniPIC, such as multi-node support with hierarchical particle decomposition through MPI, as future work.

\section*{Acknowledgment}
Funding for the work is received from the European Commission H2020 program, Grant Agreement No. 801039 (EPiGRAM-HS) and Grant Agreement No. 800904 (VESTEC). Experiments were performed on resources provided by the Swedish National Infrastructure for Computing (SNIC) at HPC2N and Lassen supercomputer at Lawrence Livermore National Laboratory. LLNL-CONF-813182.

\bibliographystyle{IEEEtran}
\bibliography{main}

\end{document}